\documentclass[
 reprint,
 superscriptaddress,
 amsmath,amssymb,
 aps,
 pra,
]{revtex4-2}

\usepackage{graphicx}
\usepackage{bm}
\usepackage{amsmath}
\usepackage{xcolor}
\usepackage{algorithm}
\usepackage{algpseudocode}

\begin{document}

\newcommand{\LR}[2][purple]{\textcolor{#1}{#2}}
\newcommand{\BO}[1]{\textcolor{blue}{#1}}

\title{Optimal multi-spectral squeezing via deterministic 2D-phase optimization}

\author{Bastien Oriot}
\thanks{These authors contributed equally to this work.}
\email{bastien.oriot@lkb.upmc.fr}
\affiliation{Laboratoire Kastler Brossel, Sorbonne Universit\'{e}, ENS-Universit\'{e} PSL, CNRS, Coll\`{e}ge de France, 4 place Jussieu, Paris F-75252, France}

\author{Peter Namdar}
\thanks{These authors contributed equally to this work.}
\affiliation{Laboratoire Kastler Brossel, Sorbonne Universit\'{e}, ENS-Universit\'{e} PSL, CNRS, Coll\`{e}ge de France, 4 place Jussieu, Paris F-75252, France}

\author{\'Emilie Gillet}
\affiliation{Laboratoire Kastler Brossel, Sorbonne Universit\'{e}, ENS-Universit\'{e} PSL, CNRS, Coll\`{e}ge de France, 4 place Jussieu, Paris F-75252, France}

\author{RL Rinc\'{o}n Celis}
\affiliation{Laboratoire Kastler Brossel, Sorbonne Universit\'{e}, ENS-Universit\'{e} PSL, CNRS, Coll\`{e}ge de France, 4 place Jussieu, Paris F-75252, France}

\author{Valentina Parigi}
\affiliation{Laboratoire Kastler Brossel, Sorbonne Universit\'{e}, ENS-Universit\'{e} PSL, CNRS, Coll\`{e}ge de France, 4 place Jussieu, Paris F-75252, France}

\date{\today}

\begin{abstract}
Optimization routines are ubiquitous in quantum information technologies and essential to reach the resource levels required by quantum protocols. Specifically, multi-spectral squeezing for use in such protocols requires that losses be kept minimal at every stage, including coherent detection, which is performed by interfering the signal with a classical local-oscillator beam. This in turn requires control over all optical degrees of freedom of the beam in order to optimize the detection.
The most general framework for this optimization relies on agnostic, off-the-shelf machine-learning techniques. Here we take the opposite approach: by focusing on a physical description of the specific optical process, we develop a deterministic sequential algorithm that provably reaches the global maximum of the visibility in a pixel basis and scales linearly with the number of pixels, thereby offering an efficient and theoretically grounded alternative to black-box optimization.
In our waveguide-based setup, the optimized mask increases the visibility from 76\% to 84\%, corresponding to a 20\% gain in mode-matching efficiency. Multi-spectral squeezing measurements confirm that this improvement translates directly into quantum readout: for the most squeezed spectral mode, the squeezing increases from $-2.08$\,dB to $-2.64$\,dB, consistent with the inferred efficiency gain. These results establish deterministic spatial phase shaping as an effective, interpretable route to enhanced multimode squeezing in waveguide platforms.
\end{abstract}

\maketitle

\section{Introduction}
Photonic platforms are among the most promising candidates for scalable quantum
computing. In the continuous-variable (CV) model, large cluster states, which are key
resources for measurement-based quantum computing (MBQC), can be generated
deterministically from multimode squeezed light combined with linear mode
transformations~\cite{Menicucci_2006,Chen14,Cai17,Yokoyama13,Asavanant_19,Larsen_2019}.
To date, the highest measured  squeezing levels have been achieved with optical parametric oscillators (OPOs), in which a $\chi^{(2)}$ nonlinear crystal is pumped inside an optical cavity~\cite{Vahlbruch_2016,Shajilal2022}. By contrast, waveguide-based systems offer broad
bandwidth \cite{Kashiwazaki2023} and natural compatibility with pulsed operation, making them particularly
attractive for time/frequency multiplexing ~\cite{Kouadou2023, Roman_Rodriguez_2024} and large-scale resource-state
generation.

In the CV approach, the encoding and decoding of information require measuring the
quadratures of quantum states. The most common technique used for this purpose is
homodyne detection, which measures the field quadratures in the spatiotemporal mode
defined by the local oscillator (LO) interfering with the signal of interest in this
coherent detection scheme. This requires a perfect overlap between the mode of the LO
and the signal, a condition that can be experimentally challenging. For instance, in an
OPO, the spatial mode of the signal is determined by the cavity geometry and is therefore well defined, enabling high mode overlap. In contrast, in a waveguide-based
system, the spatial mode may contain irregularities due to fabrication defects and
inhomogeneous waveguide geometry. Hence, mode overlap between the LO and the quantum
state becomes more difficult, limiting the performance of the homodyne detection.

While the use of a 4-$f$ pulse
shaper~\cite{FROEHLY198363,monmayrant:hal-00569786} enables full control of the
temporal mode of the local oscillator, making near-perfect temporal matching possible,
a common strategy to improve spatial overlap is to send the LO through a second
waveguide. This shapes the LO to better match the signal spatial
mode~\cite{Roman_Rodriguez_2024,kashiwazaki2020continuous}, without disturbing the
temporal mode.

More recently, approaches based on the use of a spatial light modulator (SLM) to
tailor the spatial profile of the local oscillator have been
proposed~\cite{Amari:23,ha2026generation12dbsqueezed}. Such approaches either optimize
the visibility of the interference fringes between the LO and a proxy for the quantum
signal~\cite{Amari:23}, or directly optimize the measured
squeezing~\cite{ha2026generation12dbsqueezed}. In any case, they are based on
machine-learning techniques, which are effective but not always reproducible strategies.
Moreover, they often provide limited physical insight into the optimized mode, and they
do not always guarantee convergence to the global maximum of the visibility.

In this work, we propose and implement a sequential deterministic algorithm to optimize
the mode overlap between the LO and the signal beam using an SLM, leading to higher
measured squeezing levels in different frequency modes. The method relies on: (i) the direct representation, in a pixel basis, of the SLM mask onto which the mode of interest is imaged; (ii) the use of the visibility between a proxy seed beam and the local oscillator as an objective function, which can advantageously be expressed as a sum of contributions from individual pixels; (iii) the sequential optimization of each of these contributions, that deterministically guarantees convergence to a global maximum in a reduced number of steps.

In the future, these adaptive optics approaches can be used directly to shape the mode of
the quantum signal light~\cite{doi:10.1126/science.adk7825}, optimizing the quantum feature of interest in the detection, e.g. squeezing.

\section{Deterministic optimization of the LO spatial profile}

\subsection{Imperfect mode matching}

\subsubsection*{\textbf{Effect of losses}}
The non-unit overall efficiency of a homodyne detection setup ($\eta<1$) can be
modeled as an ideal detector preceded by a loss
channel~\cite{LEONHARDT199589}. Within this framework, the observed quadrature
variance $\Delta X^2_{\mathrm{obs}}$ is related to the intrinsic variance of the input
state $\Delta X^2_{\mathrm{state}}$ through
\begin{equation}\label{loss and squeezing}
    \Delta X^2_{\mathrm{obs}}=\eta\,\Delta X^2_{\mathrm{state}} + (1-\eta),
\end{equation}
where the vacuum variance is normalized to unity.

For strongly squeezed states ($\Delta X^2_{\mathrm{state}} \ll 1$), the measured
variance approaches $\Delta X^2_{\mathrm{obs}} \approx 1-\eta$. This shows that
detection losses impose a fundamental limit on the observable squeezing, making high
overall efficiency essential for measuring large squeezing levels.

The total efficiency can be decomposed as
$\eta=\eta_{\mathrm{opt}}\,\eta_{\mathrm{mod}}\,\eta_{\mathrm{hom}}$, where
$\eta_{\mathrm{opt}}$ accounts for propagation and optical component losses,
$\eta_{\mathrm{mod}}$ describes the imperfect mode overlap between the LO and the
signal, and $\eta_{\mathrm{hom}}$ includes the finite quantum efficiency of the
photodiodes as well as the electronic efficiency of the homodyne
detector~\cite{kouadou2025homodynedetectionpulsebypulsesqueezing}. The present work
specifically aims at improving $\eta_{\mathrm{mod}}$.

\subsubsection*{\textbf{Measuring $\boldsymbol\eta_{\mathrm{mod}}$}}
A direct measurement of the overlap between the local oscillator and the quantum state
is not practically feasible, since the generated quantum field---a squeezed vacuum---is
too weak. We therefore use a standard approach: a weak coherent seed acts as a linear
probe of the detection mode; under identical propagation and polarization conditions,
it provides a reliable proxy for optimizing the overlap with the measured quantum
field~\cite{Lvovsky_2009,Aichele2002,Grosshans2001}.

The overlap between the seed beam and the local oscillator (LO) can be measured
experimentally by interfering the two beams on a beam splitter and recording the
maximum and minimum of intensity $I_\mathrm{max}$ and $I_\mathrm{min}$ at one output port as the relative phase is varied. The visibility is
then defined as
\begin{equation}
  \mathcal{V}=\frac{I_{\mathrm{max}}-I_{\mathrm{min}}}{I_{\mathrm{max}}+I_{\mathrm{min}}}.
\end{equation}
When the two beams have the same intensity, the visibility is equal to the mode overlap
between them~\cite{RevModPhys.37.231}. The visibility can be used to estimate the mode
matching efficiency according to the formula $\eta_{\mathrm{mod}}=\mathcal{V}^2$.

\begin{figure}[ht!]
\centering
\includegraphics[width=0.5\linewidth]{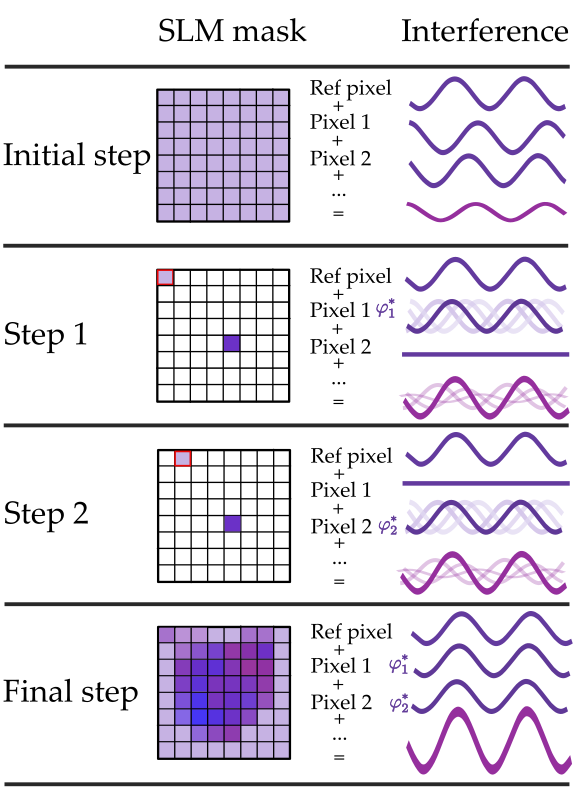}
\caption{Visual representation of the algorithm. Only the optimization of the first two
pixels is shown. The purple pixel in the middle is the reference pixel; the white pixels
are the ones reflected out and filtered in the Fourier plane. On the right is
represented the contribution of each pixel to the interference signal while the phase
between the LO and the seed is scanned.}
\label{schema_algorithm}
\end{figure}

\subsection{Visibility in the pixel basis}
As will be shown in the description of the experimental setup, the spatial mode profile of the LO is shaped by imaging it onto the SLM mask. It therefore appears natural to use the pixelized basis corresponding to square subsections of the SLM liquid-crystal array. Moreover, the elements of such a basis, besides being orthogonal, are uniquely localized in distinct spatial regions with no overlap, which, as we will see, simplifies the optimization routine.

If we denote by $f_{\mathrm{seed}}$ and $f_{\mathrm{LO}}$ the transverse spatial modes
of the two beams, their overlap can be expanded in the pixel basis on the SLM plane,
$\{p_i\}_{0 \leq i < n}$  as $f_{\mathrm{LO,seed}}=\sum_i \langle p_i, f_{\mathrm{LO,seed}}\rangle p_i $. In practice, these ``pixels'' are macropixels (e.g., a
$n = 10\times10$ partition of the active SLM area), each assigned a single
phase parameter, rather than the $1272\times1024$ native SLM pixels. The visibility
then reads:
\begin{equation}
    \mathcal{V}=|\langle f_{\mathrm{seed}}, f_{\mathrm{LO}}\rangle|
    =\bigg|\sum_i \langle f_{\mathrm{seed}}, p_i\rangle\langle p_i,f_{\mathrm{LO}}\rangle\bigg|.
\end{equation}

This expression shows that the visibility can be decomposed into contributions from
individual pixels. However, these contributions are not necessarily phase-aligned,
i.e., the phase of the complex number $ z_i=\langle f_{\mathrm{seed}},
p_i\rangle\langle p_i, f_{\mathrm{LO}}\rangle$ is not constant across pixels. Moreover,  this phase is generally not known experimentally and depends on the wavefront
distortions.

If we shape the phase of the LO by applying a phase $\varphi_i$ on pixel $p_i$ on
the SLM screen, the mode of the LO becomes
$f'_{\mathrm{LO}}=\sum_i \langle p_i, f_{\mathrm{LO}}\rangle e^{i\varphi_i} p_i$,
and the visibility reads
\begin{equation}\label{visbility with phi}
    \mathcal{V}(\varphi_0,\ldots,\varphi_n)
    =\bigg|\sum_i \langle f_{\mathrm{seed}}, p_i\rangle\langle p_i, f_{\mathrm{LO}}\rangle
    e^{i\varphi_i}\bigg|.
\end{equation}
This shows that by properly tuning the phases $\varphi_i$ the pixel-wise contributions
can be made to sum constructively. Choosing
$\varphi^*_i=-\arg(z_i)$ all the elements in Eq.~(\ref{visbility with phi}) sum up in phase, i.e. the result is the sum of absolute values of each term yielding to the maximal value of the visibility that is  achievable with phase-only shaping:
\begin{equation}\label{visam}
    \mathcal{V}(\varphi_0^*,\ldots,\varphi_n^*)
    =\sum_i|\langle f_{\mathrm{seed}}, p_i\rangle\langle p_i, f_{\mathrm{LO}}\rangle|.
\end{equation}

With phase-only shaping, the visibility maximum remains below 1. Adding amplitude
shaping can enable perfect overlap by matching the pixel-wise powers
of the seed and LO fields.

\subsection{Sequential algorithm to design the best spatial phase mask}

The goal is to determine the optimal phase map $\varphi_0^*,\ldots,\varphi_{n - 1}^*$ that
maximizes $\mathcal{V}(\varphi_0,\ldots,\varphi_{n - 1})$. A naive random search over
all parameters does not, in general, guarantee a global optimum. As we have to align the phase of the different terms in Eq.~(\ref{visbility with phi}) to get the optimized Eq.~(\ref{visam}) we decided to use a
reference pixel $p_0$ as a phase anchor and optimize the remaining pixels one by one, sequentially.

This approach is possible only because the fitness function we optimize, the visibility, is the sum of the contributions from each individual pixel and does not depend on cross-terms, that is, on pairs of pixels.

For each pixel $p_i$, we prepare a two-pixel mode of the form $p_0+e^{i\varphi}p_i$
and scan the relative phase $\varphi$ over $[0,2\pi)$ while measuring the current
visibility:
\begin{equation}\label{visbilite_pixel}
    \mathcal{V}_i(\varphi)
    = |\langle f_{\mathrm{seed}}, p_0\rangle\langle p_0,f_{\mathrm{LO}}\rangle
      +e^{i\varphi}\langle f_{\mathrm{seed}},p_i\rangle\langle p_i, f_{\mathrm{LO}}\rangle|.
\end{equation}
For each $i$, the phase $\varphi_i$ that maximizes $\mathcal{V}_i(\varphi)$ is
retained, and the procedure is repeated for all pixels. The full algorithm is
summarized in Fig.~\ref{alg:sequential_phase_optimization}. Note that since
$\mathcal{V}_i(\varphi)$ varies sinusoidally with $\varphi$, the optimum can be obtained from a least-square sinusoidal fit to a small number of sampled points rather than from a complete scan. In practice, $p_0$ is chosen at the center of the SLM screen, where the beam intensity is the highest, to measure a strong and reliable visibility signal.

The key element here is that by maximizing $\mathcal{V}_i(\varphi)$, we align the
phase of each pixel $p_i$ with the common reference phase $p_0$. This alignment
ensures that all contributions in the full sum of Eq.~(\ref{visbility with phi})
point in the same direction and interfere constructively, thereby maximizing the overall
visibility as illustrated in Fig.~\ref{schema_algorithm}. We thus reduced this
$n$-parameter optimization problem to $n$ single-parameter optimization problems.

\begin{figure}[ht]
\centering
\includegraphics[width=0.5\linewidth]{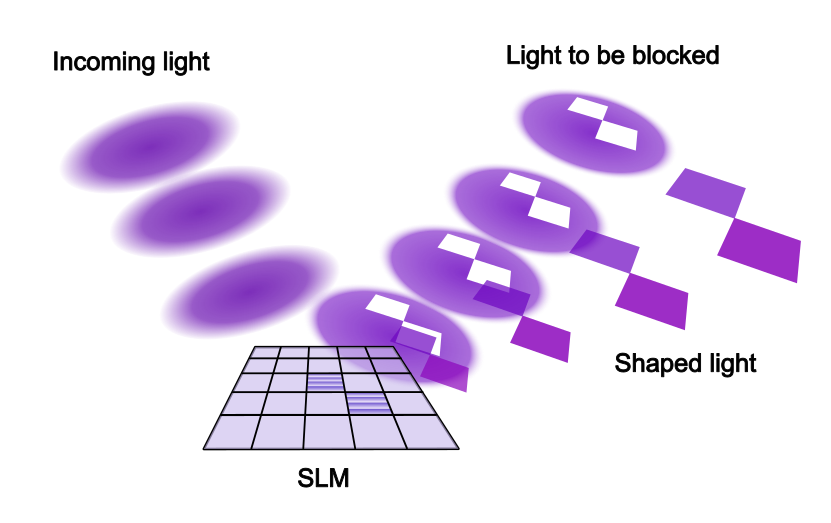}
\caption{Zoom on the SLM: on the SLM, the phase ramp is applied only on
two pixels. The rest of the light is thus reflected in another direction and can be
blocked in the Fourier plane.}
\label{schema_phase_ramp}
\end{figure}

Experimentally, measuring $\mathcal{V}_i$ requires isolating only the two target pixels
$p_0$ and $p_i$ while suppressing all others. To achieve this, we apply a linear phase
ramp to the selected pixels and no ramp to the rejected ones
(Fig.~\ref{schema_phase_ramp}). The phase ramp creates spatial separation in the
Fourier plane between selected and rejected contributions, allowing the desired mode to
be isolated with an iris.

\begin{figure}[t]
\centering
\caption{Algorithm of the sequential phase optimization with reference pixel}
\label{alg:sequential_phase_optimization}
\begin{algorithmic}
\Require Pixel basis $\{p_i\}_{0 \leq i < n}$, reference pixel $p_0$
\Ensure Optimal phase map $\{\varphi_i^*\}_{0 \leq i < n}$
\State Set $\varphi_0^* \gets 0$ \Comment{Phase reference}
\For{$i=1$ to $n - 1$}
    \For{$\varphi$ in $[0,2\pi)$ with step $\delta\phi$}
        \State Send the mask: $p_0+e^{i\varphi}p_i$
        \State Measure visibility $\mathcal{V}_i(\varphi)$
    \EndFor
    \State $\varphi_i^*\gets \text{argmax}\bigl(\mathcal{V}_i(\varphi)\bigr)$
\EndFor
\State \Return $\{\varphi_i^*\}_{0 \leq i < n}$
\end{algorithmic}
\end{figure}


\section{Experimental set-up}
\begin{figure}[ht]
\centering
\includegraphics[width=0.98\linewidth]{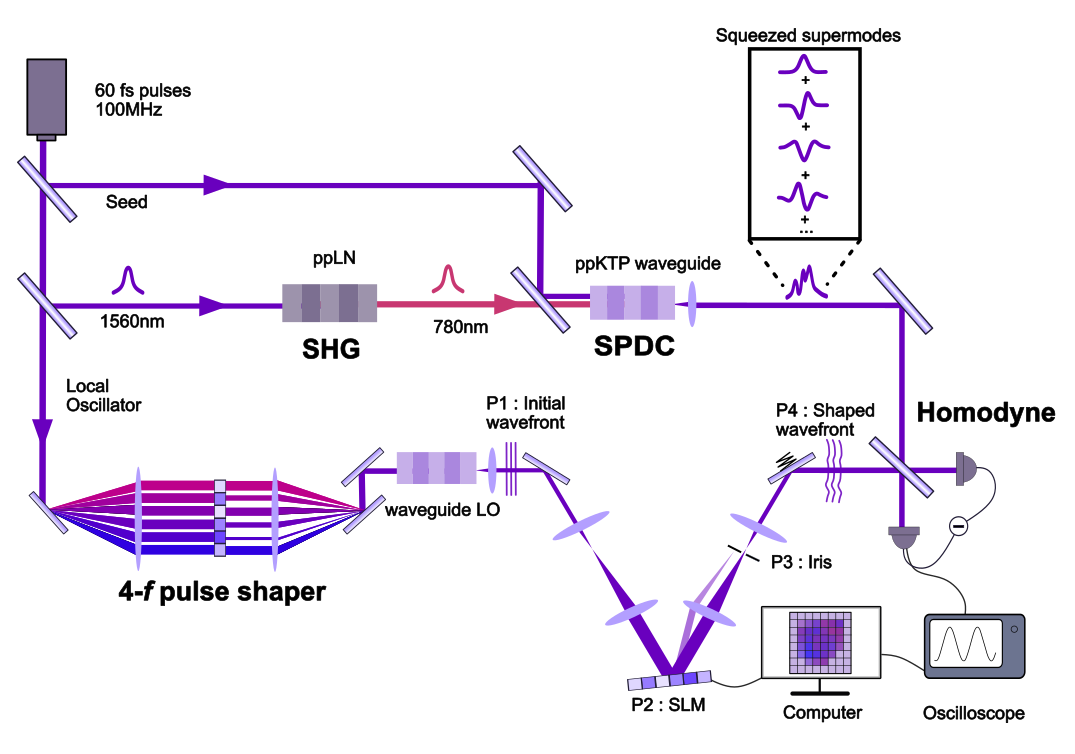}
\caption{Scheme of the experimental setup. The second harmonic of the femtosecond laser pumps a ppKTP crystal for squeezing generation. A second portion of the main laser is spectrally shaped and spatially optimized to serve as the local oscillator in homodyne detection. When squeezing is measured, the seed path is blocked. For spatial phase optimization of the LO, the 780\,nm beam is blocked while the seed is used to mimic the quantum signal. To maximize its overlap with the LO, a photodiode is placed at the beam splitter (BS) output to monitor the interference fringes.}
\label{experimental_setup}
\end{figure}

\subsection*{Squeezing measurement}

The experimental setup shown in Fig.~\ref{experimental_setup} uses a pulsed femtosecond laser operating at telecom wavelengths to generate multi-spectral squeezed light. The second harmonic at 780 nm pumps a periodically poled KTP (ppKTP) waveguide, producing a spectrally multimode squeezed state that is subsequently analyzed via homodyne detection \cite{Roman_Rodriguez_2024,kouadou2025homodynedetectionpulsebypulsesqueezing}.  Owing to the interplay between the waveguide nonlinearity and the finite spectral bandwidth of the pump, the generated down-converted light exhibits multimode spectral squeezing. This structure can be described in terms of a set of orthogonal squeezed modes, known as supermodes, each characterized by a squeezing parameter and an associated spectral profile \cite{Patera_2009}. Numerical simulations indicate that these mode functions are well approximated by the Hermite-Gauss basis, whose elements are denoted by $HG_n$, and which can be selectively addressed through spectrally mode-selective homodyne detection.
To this end, the local oscillator (LO) is spectrally shaped using a 4-$f$ pulse shaper and subsequently coupled into a second waveguide to approximately reproduce the spatial intensity profile of the signal. By appropriately choosing the polarization, nonlinear interactions within this waveguide are suppressed.

Finally, the LO is refined with a second SLM for spatial phase-profile mode matching. This last step constitutes the new module of the experimental setup. Once optimized, as explained in the following section, the shaped LO is used in homodyne detection for squeezing measurements.
The relative phase between the signal and the local oscillator is scanned using a piezo-actuated mirror.

\subsection*{Visibility optimization}

To determine the optimal phase mask for squeezing measurements, we use a modified
experimental configuration in which the pump is blocked and only the seed beam is
injected into the waveguide. The seed acts as a proxy for optimizing the visibility \cite{Aichele2002,Grosshans2001}. A
photodiode placed after the homodyne beam splitter monitors  the interference fringes  between the seed and local oscillator while their relative phase is scanned with a
piezoelectric mirror. The visibility is measured and computed on an oscilloscope
(LeCroy WaveRunner 8108HD) and sent to the computer controlling the SLM (Hamamatsu
X15213). The SLM active area measures $15.9\times12.8$\,mm with $1272\times1024$ pixels.

Our aim is to tailor the phase of the local oscillator at the waveguide output lens in plane \,P1 of Fig.~\ref{experimental_setup} to match the aberrated phase at the output of the waveguide generating the quantum signal. To this end, P1 is imaged onto the SLM
using a two-lens telescope with magnification 2, ensuring efficient use of the SLM
active area. The typical beam diameter on the SLM is approximately 6\,mm. To implement
the sequential algorithm, we address only two target pixels at a time with a controlled
relative phase, while all other SLM regions are suppressed by removing the grating in
those areas. This deflects the unwanted light away from the desired mode, which is then
isolated using an iris in plane\,P3 (the Fourier plane of P2). After optimization,
plane\,P4 contains the phase-corrected beam.

\section{Results}
\begin{figure}[ht]
\centering
\includegraphics[width=0.7\linewidth]{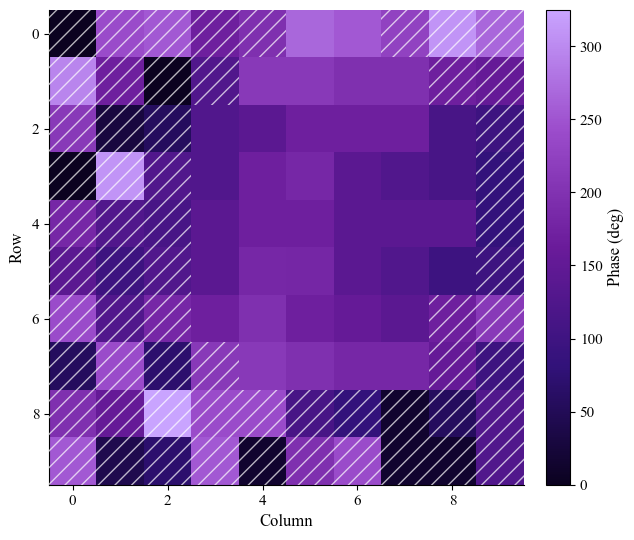}
\caption{Obtained phase map; the non-hatched portion shows where the beam
is physically on the screen.}
\label{mask}
\end{figure}

We performed several optimization runs with different SLM pixel grids. Although the
algorithm runs successfully with a $(15\times15)$ grid, a $(10\times10)$ grid already
provided nearly identical results and was therefore retained. For this grid size, the
algorithm runs in less than 5 minutes, and the runtime is dominated by the acquisition
time of a stable visibility signal on the oscilloscope for each candidate phase. This
method is highly reproducible: successive runs yield nearly identical phase masks.
Moreover, the optimized mask does not need to be recomputed on a daily basis; it
remains valid as long as the same waveguides, on the signal and the LO paths, are used. The obtained phase mask for a specific pair of signal and LO waveguides of the chips used in the experimental setup \cite{Roman_Rodriguez_2024}, is shown in Fig.~\ref{mask} and it is used for the following squeezing measurement.

\begin{figure}[ht]
\centering
\includegraphics[width=0.9\linewidth]{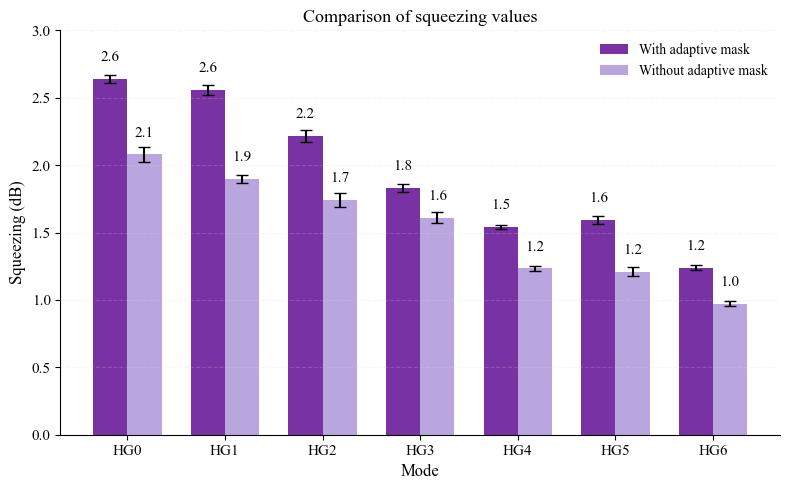}
\caption{Measurement of squeezing with and without the adaptive mask.}
\label{squeezingm}
\end{figure}

Visibility typically increased from 76\% to 84\%. The optimization thus makes the
visibility go from $\eta=0.76^2\eta_{\mathrm{opt}}\eta_{\mathrm{hom}}$ to
$\eta'=0.84^2\eta_{\mathrm{opt}}\eta_{\mathrm{hom}}$, which corresponds to a 20\%
increase in detection efficiency. To validate the impact of this visibility enhancement
on squeezing, we measured squeezing in several frequency
supermodes~\cite{Patera_2009,Roslund_2013} of our system with and without the
algorithm-derived phase mask. Importantly, no realignment of the optical setup was
performed between the two masks, ensuring that any squeezing improvement is
attributable only to phase shaping. Shot noise was remeasured for each mask, even
though it was constant, because the phase mask does not affect the optical power. Using a homodyne detector with a bandwidth of around 10\,MHz and measuring at an electronic frequency of 1\,MHz in zero-span mode with a spectrum analyzer (Agilent N9020A) we
measured quadrature variances by scanning the LO phase. For each mode, we recorded
about 15 values of the squeezed-quadrature variance, which allowed us to estimate the
statistical uncertainty. Fig.~\ref{squeezingm} shows the improvement in the measured squeezing for the first six candidate supermodes, that are Hermite-Gauss frequency modes centered at 1560 nm and with a full width at half maximum (FWHM) of 15 nm for the $\text{HG}_0$ mode. More informations can be found in~\cite{Roman_Rodriguez_2024,kouadou2025homodynedetectionpulsebypulsesqueezing}.

For the $\text{HG}_0$ mode, squeezing improved from
$-2.08\pm0.03$\,dB to $-2.64\pm0.02$\,dB, corresponding to an inferred efficiency
increase of $(20\pm3)\,\%$ according to Eq.~(\ref{loss and squeezing}), in very good
agreement with the value expected from the visibility change. This confirms that the
seed beam is a relevant proxy for the quantum signal to be measured. Across the other modes, mean efficiency rose by 21\%; the observed fluctuations, which are most pronounced in the $\text{HG}_3$ mode, are likely attributable to variations in the spatial overlap occurring during mode switching.

\section{Conclusion}
In this work, we developed a method to compute the optimal phase-correction mask that
compensates for the relative phase mismatch between the local oscillator (LO) mode and
the target mode to be measured. Notably the fitness function, the visibility fringes between a proxy of the quantum signal and the local oscillator, depends linearly on the parameters to be optimized, allowing for sequential optimization. As a consequence the
computational complexity of the procedure scales
linearly with the number of pixels to be optimized.

Using this approach, we improved the interference visibility from $76\%$ to $84\%$,
corresponding to an increase of approximately $20\%$ in detection efficiency. The
shaped LO was then used to measure squeezing, leading to enhanced performance in
several frequency Hermite--Gaussian modes. For the first mode, the measured squeezing
improved from $-2.08\pm0.02$\,dB to $-2.64\pm0.03$\,dB, in excellent agreement with
the expected increase in efficiency.

The origin of the remaining visibility gap is still under investigation. We estimate
that approximately 10\% of the loss can be attributed to an intensity-profile mismatch
that cannot be corrected by phase-only shaping. The residual limitation may arise from
weak parasitic reflections within the setup, which appear to generate after-pulses
reducing the interference contrast; however, we are not yet able to quantify
the contribution of this effect.

\begin{acknowledgments}
This project has received funding from the European Union's Horizon Europe Framework
Programme under Grant Agreement No.\ 101114899 (VeriQub), and by Agence Nationale de
la Recherche (OQuLus, ANR-22-PETQ-0013).
The authors thank C.\ Verni\`{e}re, B.\ Courme and H.\ Defienne for fruitful discussions.
The authors declare no conflicts of interest.
\end{acknowledgments}

\section*{Data availability}
Data underlying the results presented in this paper are not publicly available at this
time but may be obtained from the authors upon reasonable request.

\bibliography{bibliography}

\end{document}